\documentclass[12pt,showkeys,showpacs,] {revtex4}
\usepackage{graphicx}
\begin{document}

\title[Short Title]{Quantum Dialogue by Using the GHZ State}
\author{Yan \surname{XIA}}
\author{Chang-Bao \surname{FU}}
\author{Shou \surname{ZHANG}\footnote{E-mail: szhang@ybu.edu.cn} }
\affiliation{Department of Physics, College of Science, Yanbian
University, Yanji, Jilin 133002, PR China}
\author{Suc-Kyoung \surname{HONG}}
\author{Kyu-Hwang \surname{YEON}}
\affiliation{Department of Physics,  Institute for Basic  Science
Research, College of Natural Science, Chungbuk National University,
Cheonju, Chungbuk 361-763}

\author{Chung-In \surname{UM}}
\affiliation{Department of Physics, College of Science, Korea
University, Seoul 136-701}

\begin{abstract} We present a quantum dialogue protocol by using the Greenberger-Horne-Zeilinger (GHZ) state. In
this paper, we point out that the `quantum dialogue' communication
scheme recently introduced by Nguyen can be eavesdropped on under
an intercept-and-resend attack. We also give a revised control
mode to detect this attack. Hence, within the present version two
users can exchange their secret messages securely and
simultaneously, and the efficiency of information transmission can
be successfully increased.
 \end{abstract}
 \maketitle Since the pioneering work of Bennett and
Brassard published in 1984 \cite{BB84}, different quantum key
distribution protocols have been presented
\cite{ABCHPRA02,SJGPRA02,XLGPRA02,SPRA04,WPRL04}. Different from
the key distribution protocol, some quantum direct secure
communication (QDSC) protocols have been shown recently
\cite{BFPRL02,DLLPRA03,MZLCPL0518,XFZJKPS05}, which permit
important messages to be communicated directly without first
establishing a random key to encrypt them. However, these
protocols
\cite{BB84,ABCHPRA02,SJGPRA02,XLGPRA02,SPRA04,WPRL04,BFPRL02,DLLPRA03,MZLCPL0518,XFZJKPS05}
only permit messages to be transmitted from the sender (Alice) to
the receiver (Bob). Two parties cannot simultaneously transmit
their different secret messages to each other in only one quantum
channel.

Very recently,  Nguyen \cite{NBA04PLA} proposed an
entanglement-based protocol, the so-called quantum dialogue, which
would allow two people to exchange their messages simultaneously
based on the well-known Bostroem-Felbinger \cite{BFPRL02}
protocol. Man {\it et al.} \cite{MZLCPL05} proposed a modified
protocol about the quantum dialogue. It is worth mentioning that
the Bostroem-Felbinger protocol can be eavesdropped in some
specific cases \cite{W03PRL,ZMLPLA}. In this paper, we show that
the quantum dialogue scheme \cite{NBA04PLA} is insecure under some
eavesdropping attacks, and we present a new protocol to realize a
quantum dialogue by using the GHZ state more securely. Thus, the
two users can exchange their messages securely and simultaneously.
The efficiency of information transmission is also successfully
increased.

Now, let us review Nguyen's quantum dialogue scheme. Suppose there
are two users (say, Alice and Bob) and they want to transmit their
messages to each other simultaneously. Bob first produces a large
enough number of Einstein-Podolsky-Rosen (EPR) pairs, all in the
state
\begin{equation}\label{e1}
|\Psi_{0,0}\rangle_{ht}=\frac{1}{\sqrt{2}}(|\uparrow\rangle_h|\downarrow\rangle_t+|\downarrow\rangle_h|\uparrow\rangle_t),
\end{equation}
where {\it h} stands for `` {\it home}, '' {\it t} stands for ``
{\it travel}, '' and $|\downarrow\rangle$ and $|\uparrow\rangle$
characterize two degrees of freedom of a qubit. Bob encodes his
bits $(k_n, l_n)$ $(k_n, l_n\in\{0,1\})$ by applying an operation
$C_{k_n,l_n}$ on the state $|\Psi_{0,0}\rangle_{ht}$, where
$C_{0,0}$, $C_{0,1}$, $C_{1,0}$, and $C_{1,1}$ denote the Pauli
matrices $I$, $\sigma_x$, $\sigma_y$, and $\sigma_z$,
 respectively. He keeps one
qubit (home qubit) with him and sends another (travel qubit) to
Alice. Then, Bob lets Alice know that he has sent a qubit. Alice
tells Bob that she has received a qubit. Alice encodes her bits
$(i_n,j_n)$ $(i_n, j_n \in\{0,1\})$ by performing an operation
$C_{i_n,j_n}$ on the travel qubit; then, she sends it back to Bob.
When Bob receives the encoded travel qubit, he performs a Bell
basis measurement on the qubit pair and waits for Alice to tell
him that it was a run in a message mode (MM) or in a control mode
(CM). In a MM run, Bob decodes Alice's bits and announces his Bell
basis measurement result $(x_n,y_n)$ to let Alice decode his bits.
In a CM run, Alice reveals her encoding value to Bob to check the
security of their dialogue.

However, this security checking cannot detect Eve's
(eavesdropper's) intercept-and-resend attack. Let us suppose Eve
gets the qubit {\it t} and keeps it with her. Afterwards, she
creates her own entangled pair in the same state as in
Eq.~(\ref{e1}), i.e., Eve's pair state is
\begin{equation}\label{e2}
|\Psi_{0,0}\rangle_{HT}=\frac{1}{\sqrt{2}}(|\uparrow\rangle_H|\downarrow\rangle_T+|\downarrow\rangle_H|\uparrow\rangle_T),
\end{equation}
and sends her qubit {\it T} to Alice. Alice, taking {\it T} for
{\it t}, encodes her bits by performing an appropriate
transformation as described above and sends it back to Bob. Then,
Eve intercepts the qubit {\it T} again and carries out a Bell
basis measurement on the {\it {HT}}- pair to learn Alice's secret
bits. By the same Bell basis measurement, Eve knows the encoding
transformation Alice performed on the qubit {\it T}. Eve then
applies the same transformation on the qubit {\it t} she has kept
and sends it back to Bob. After Bob announces publicly his Bell
basis measurement result, Eve can deduce Bob's bits. Clearly, Eve
can eavesdrop completely on the contents of Bob and Alice's
dialogue. Even worse, Eve's tampering is absolutely unnoticeable.

Now, we propose our new protocol to realize quantum dialogue by
using the GHZ state in terms of the original scheme. However in
our protocol the intercept-and-resend attack, as well as the other
kinds of attacks presented by Nguyen \cite{NBA04PLA}, can be
detected. First, we write the eight GHZ state bases in two
different bases as follows:
\begin{eqnarray}\label{e3}
&|\psi_{0,0;0,0}\rangle&=\frac{1}{\sqrt{2}}(|000\rangle+|111\rangle)_{htp}\cr\cr&&=\frac{1}{2}[|+\rangle_h(|+\rangle_t|+\rangle_p+
|-\rangle_t|-\rangle_p)+|-\rangle_h(|+\rangle_t|-\rangle_p+|-\rangle_t|+\rangle_p)],
\end{eqnarray}
\begin{eqnarray}\label{e4}
&|\psi_{1,1;0,0}\rangle&=\frac{1}{\sqrt{2}}(|000\rangle-|111\rangle)_{htp}\cr\cr&&=\frac{1}{2}[|+\rangle_h(|-\rangle_t|+\rangle_p+
|+\rangle_t|-\rangle_p)+|-\rangle_h(|+\rangle_t|+\rangle_p+|-\rangle_t|-\rangle_p)],
\end{eqnarray}
\begin{eqnarray}\label{e5}
&|\psi_{0,1;0,0}\rangle&=\frac{1}{\sqrt{2}}(|100\rangle+|011\rangle)_{htp}\cr\cr&&=\frac{1}{2}[|+\rangle_h(|+\rangle_t|+\rangle_p+
|-\rangle_t|-\rangle_p)-|-\rangle_h(|+\rangle_t|-\rangle_p+|-\rangle_t|+\rangle_p)],
\end{eqnarray}
\begin{eqnarray}\label{e6}
&|\psi_{1,0;0,0}\rangle&=\frac{1}{\sqrt{2}}(|100\rangle-|011\rangle)_{htp}\cr\cr&&=\frac{1}{2}[|+\rangle_h(|+\rangle_t|-\rangle_p+
|-\rangle_t|+\rangle_p)-|-\rangle_h(|+\rangle_t|+\rangle_p+|-\rangle_t|-\rangle_p)],
\end{eqnarray}
\begin{eqnarray}\label{e7}
&|\psi_{0,0;0,1}\rangle&=\frac{1}{\sqrt{2}}(|010\rangle+|101\rangle)_{htp}\cr\cr&&=\frac{1}{2}[|+\rangle_h(|+\rangle_t|+\rangle_p-
|-\rangle_t|-\rangle_p)+|-\rangle_h(|+\rangle_t|-\rangle_p-|-\rangle_t|+\rangle_p)],
\end{eqnarray}
\begin{eqnarray}\label{e8}
&|\psi_{1,1;0,1}\rangle&=\frac{1}{\sqrt{2}}(|010\rangle-|101\rangle)_{htp}\cr\cr&&=\frac{1}{2}[|+\rangle_h(|+\rangle_t|-\rangle_p-
|-\rangle_t|+\rangle_p)+|-\rangle_h(|+\rangle_t|+\rangle_p-|-\rangle_t|-\rangle_p)],
\end{eqnarray}
\begin{eqnarray}\label{e9}
&|\psi_{0,1;0,1}\rangle&=\frac{1}{\sqrt{2}}(|110\rangle+|001\rangle)_{htp}\cr\cr&&=\frac{1}{2}[|+\rangle_h(|+\rangle_t|+\rangle_p-
|-\rangle_t|-\rangle_p)+|-\rangle_h(|-\rangle_t|+\rangle_p-|+\rangle_t|-\rangle_p)],
\end{eqnarray}
\begin{eqnarray}\label{e10}
&|\psi_{1,0;0,1}\rangle&=\frac{1}{\sqrt{2}}(|110\rangle-|001\rangle)_{htp}\cr\cr&&=\frac{1}{2}[|+\rangle_h(|+\rangle_t|-\rangle_p-
|-\rangle_t|+\rangle_p)+|-\rangle_h(|-\rangle_t|-\rangle_p-|+\rangle_t|+\rangle_p)],
\end{eqnarray}
where
\begin{equation}\label{e11}
|+\rangle=\frac{1}{\sqrt{2}}\ (|0\rangle+|1\rangle),
\end{equation}
\begin{equation}\label{e12}
|-\rangle=\frac{1}{\sqrt{2}}\ (|0\rangle-|1\rangle),
\end{equation}
{\it h} stands for `` {\it home}, '' {\it t} stands for `` {\it
travel}, '' and {\it p} stands for `` {\it post} ''.

 Suppose that Alice has a secret message consisting of $4N$
bits; i.e., Alice's message $=\{(i_1, j_1, f_1, g_1), (i_2, j_2,
f_2, g_2),\ .\ .\ .\ ,(i_N, j_N, f_N, g_N) \}$, where $i_n, j_n,
g_n \in \{0, 1\}$, $f_n$=$0$. Bob has another secret message
consisting of $4N$ bits, too; i.e., Bob's message $=\{(k_1, l_1,
w_1, v_1), (k_2, l_2, w_2, v_2),\ .\ .\ .\ , (k_M, l_M, w_M,
v_M)\}$, where $k_n, l_n, v_n
 \in \{0, 1\}$, $w_n$=$0$. For Bob and Alice, each of them can encode their own
 secret message (a, b, c, d) ($a, b, d \in [0, 1],c=0$) by
 performing the unitary operation $C_{a,b}^t \otimes C_{c,d}^p$ on the
 travelled qubits; $|\Psi_{a, b; c, d}\rangle=C_{a,b}^t \otimes
 C_{c,d}^p
 |\psi_{0,0;0,0}\rangle$,, where $C_{0,0}^t$, $C_{0,1}^t$, $C_{1,0}^t$, $C_{1,1}^t$, $C_{0,0}^p$, and $C_{0,1}^p$ denote $I^t$,
$\sigma_x^t$, $\sigma_y^t$, $\sigma_z^t$, $I^p$, and $\sigma_x^p$,
 respectively.

 (S1) To securely exchange their messages or, in other words, to carry out
 a secret dialogue, Bob first produces a large enough number of
 GHZ states, all in the state $|\psi_{0,0;0,0}\rangle$.  Bob keeps particles {\it $h$}
 with him and chooses particles ({\it $t$}, {\it
 $p$}) as encoding-decoding group particles.
  Bob and Alice arrange to only perform one of the eight operations on particles ({\it $h$}, {\it $p$}) as
 \begin{equation}\label{e13}
 \begin{array}{cccc}
 U_0=I^t \otimes I^p,
 \ \ U_1=\sigma_z^t \otimes I^p,
 \ \ U_2=\sigma_x^t \otimes I^p,
 \ \ U_3=i \sigma_y^t \otimes I^p,\cr\\
 U_4=I^t \otimes \sigma_x^p,
 \ \ U_5=\sigma_z^t \otimes \sigma_x^p,
 \ \ U_6=\sigma_x^t \otimes \sigma_x^p,
 \ \ U_7=i \sigma_y^t \otimes \sigma_x^p.
 \end{array}
 \end{equation}

 (S2) Bob encodes his bits $(k_n, l_n,w_n, v_n$) by
applying $U_B=C_{k_n,l_n}^t \otimes C_{w_n, v_n}^p$$(B \in
\{0,1,2,3,4,5,6,7\})$ on the GHZ state $|\psi_{0,0;0,0}\rangle$.
He keeps qubits {\it $h$} with him and sends qubits ({\it ${t}$},
{\it $p$}) through two different quantum channels to Alice. Then,
Bob lets Alice know that he has sent qubits ({\it ${t}$}, {\it
$p$}).

(S3) Alice confirms to Bob that she has received the qubits. Bob
tells Alice which mode they are: the message mode (MM) run or the
control mode (CM) run. If it is a CM (MM) run, the procedure goes
to S4 (S5).

(S4) Alice selects randomly sufficiently large groups as checking
groups and the groups leftover as encoding-decoding groups. She
chooses randomly one of the two sets of measuring basis (MB),
$\{|0\rangle, |1\rangle\}$ or $\{|+\rangle, |-\rangle\}$, by
measuring her qubits to check for the intercept-and-resend attack.
She announces her choices and her measurement outcomes. Bob
performs his measurement under the same MB as that chosen by Alice
on the corresponding photons in his checking groups. If their
measurement outcomes coincide when both of them use the same basis
according to Eq.~{(\ref{e3})$-$(\ref{e10})}. There are no Eves in
line. This communication continues. In this case, Alice and Bob
continue to transmit the next bits. Otherwise, they have to
discard their transmission and abort the communication.

(S5) Alice encodes her bits $(i_n, j_n, f_n, g_n$) by applying
$U_A$=$C_{i_n,j_n}^t$$\otimes$ $C_{f_n, g_n}^p$$(A \in\{
0,1,2,3,4,5,6,7\})$ on the encoding-decoding groups. Then, Alice
sends back qubits ({\it ${t}$}, {\it $p$}) to Bob, and lets Bob
know that.

(S6) Aware of Alice's confirmation, Bob performs a GHZ state basis
measurement on the three qubits. The result in state
$|\Psi_{(x_n,y_n);(r_n, s_n)}\rangle$ $(x_n,y_n,r_n,s_n\in
\{0,1\})$ is
\begin{eqnarray}\label{e14}
&|\Psi_{(x_n,y_n);(r_n,s_n)}\rangle&=U_BU_A|\psi_{0,0;0,0}\rangle\cr\cr&&=
C_{k_n,l_n}^t \otimes C_{w_n, v_n}^p \otimes C_{i_n,j_n}^t\otimes
C_{f_n, g_n}^p |\psi_{0,0;0,0}\rangle\cr\cr&&= C_{i_n \oplus
k_n,j_n \oplus l_n}^t \otimes C_{f_n \oplus w_n,g_n \oplus
v_n}^p|\psi_{0,0;0,0}\rangle\cr\cr&&
=\phi_{i_n,j_n;k_n,l_n}\phi_{f_n,g_n;w_n,v_n}|\psi_{i_n \oplus
k_n, j_n \oplus l_n; f_n \oplus w_n, g_n \oplus v_n}\rangle,
\end{eqnarray}
where $\oplus$ denotes an addition mod 2, $\phi_{i,j;k,l}$ and
$\phi_{f,g;w,v}$ are phase factors ($\phi=1$ or $\pm i$ depending
on the values of ($i,j,k,l$) and ($f,g,w,v$)).

(S7) With a certain possibility, Alice and Bob reveal some bits to
check the entangle-and-measure attack when particles are
travelling from Alice to Bob. If there are no attacks, Alice's
bits and Bob's bits should have the deterministic correlation
$(i_n=\mid x_n-k_n\mid$, $j_n=\mid y_n-l_n\mid$, $f_n=\mid r_n-w_n
\mid$, $g_n=\mid s_n-v_n\mid)$. The communication continues.
Otherwise, Bob publicly tells Alice that Eve is in the line, and
the communication is aborted.

(S8) Bob decodes Alice's secret bits as $(i_n=\mid x_n-k_n\mid$,
$j_n=\mid y_n-l_n\mid$, $f_n=\mid r_n-w_n \mid$, $g_n=\mid
s_n-v_n\mid)$, and he reads the 3 bits of information. If he
publicly announces the values of $(x_n, y_n, r_n, s_n)$ , then
Alice can also decode Bob's secret bits as $(k_n=\mid
x_n-i_n\mid$, $l_n=\mid y_n-j_n\mid$, $w_n=\mid r_n-f_n\mid$,
$v_n=\mid s_n-g_n\mid)$, and she read the 3 bits of information,
too. Then the procedure goes to S2, and Bob and Alice continue to
transmit their next bits. The dialogue has been successfully
completed.

In some cases, Alice and Bob can take some bits as checking bits,
and they can reveal the value of the checking bits to check for
the entangle-and-measure attack \cite{NBA04PLA} when the two
qubits are travelling from Alice to Bob; i.e., Bob takes $(k_n,
l_n)$ as secret bits and $(w_n, v_n)$ as checking bits. Alice
takes $(i_n, j_n)$ as secret bits and $(f_n, g_n)$ as checking
bits, and publicly reveals the value of the checking bits. Then,
Bob checks for Eves: if both $f_n= \mid r_n-w_n\mid $ and
$g_n=\mid s_n-v_n\mid$, Bob holds; there are no Eves measuring the
two qubits ($t_n$, $p_n$). The communication is secure. Otherwise,
Bob and Alice have to discard their transmission and abort the
communication.

In conclusion, we have proposed a new quantum dialogue protocol
for two legitimate parties to exchange their secret messages
simultaneously by using the GHZ state. The result shows that, for
such a GHZ state quantum channel, a quantum dialogue can be
realized between two users successfully. Compared with previous
schemes, there are some differences in scheme proposed in this
paper. Firstly, the quantum channel is different. Secondly,
because the encoding-decoding particles are sent with a large
number of checking particles simultaneously through two different
quantum channels, it is hard for Eve to catch the right two
particles in one group of the encoding-decoding groups
simultaneously. Thus, the total probability of an eavesdrop by Eve
is very small. This scheme can also be generalized to the
$N$-particle GHZ state system; Bob keeps particle 1 with him in
each $N$-particle GHZ state and sends the remaining $N-1$
particles to Alice. Thus, $N-1$ channels are needed, so the total
probability of an eavesdrop by Eve will go down with increasing
number of channels. Thirdly, Alice and Bob can take some bits as
checking bits and reveal the value of the checking bits to check
for the entangle-and-measure attack when the qubits are sent from
Alice to Bob, making sure the communication is secure. Fourthly,
the protocol makes subtle use of superdense coding
\cite{BWPRL92,YWCRL04,FSLZKBRS05} to double the quantum channel,
and each party is able at the same time to send three secret bits
information and to read three secret bits information. The
efficiency of information transmission is successfully increased.
The intercept-and-resend attack, as well as the
entangle-and-measure attack, can be detected efficiently in our
protocol. Though it is experimentally more difficult to prepare
GHZ states as compared to EPR states, for such a GHZ state quantum
channel, we can not only increase the effective information but
also improve the security in our protocol, so our protocol is
feasible.

\end{document}